# The possibility of solving a 3x3 Rubik's Cube under 2 seconds - Optimizing block building

Dr. KONG, Chung To

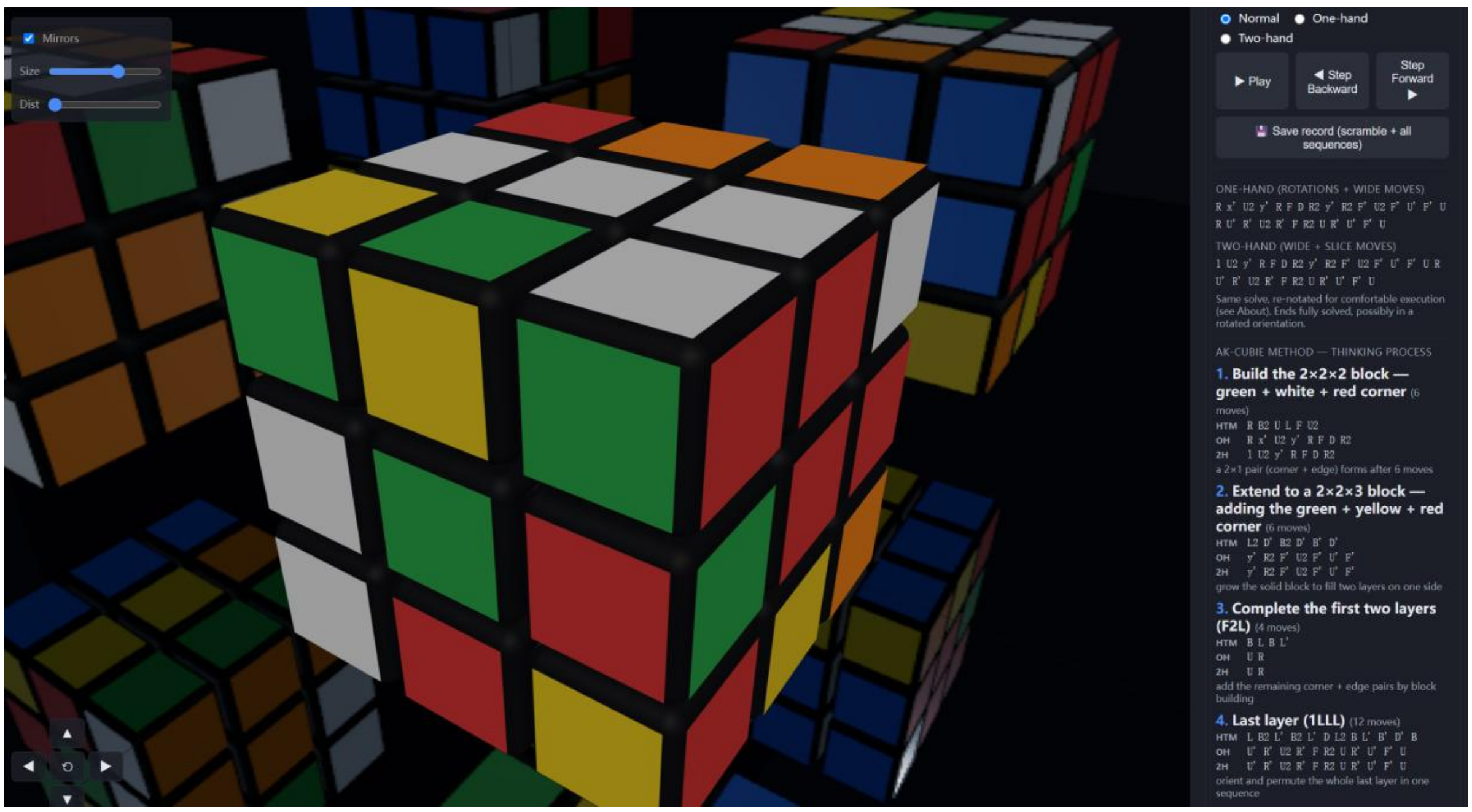

*Fig 1. AK-cubie method User Interface*

*Abstract*— Block building is an essential skill in speedcubing competitions. However, there is currently no systematic method to understand or analyze block-building techniques. Rewatching gameplay reveals only the outcome, not the underlying mechanism. This paper presents a systematic approach, the "AK-cubie method" (or "cubie method" for short), for constructing near-optimal block-building move sequences within a short computation time. Speedcubers can use the accompanying software to train their look-ahead and block-building skills. We hope this work will contribute to breaking the 2-second world record in the near future.

## I. Introduction

Every competition is defined by a set of limiting factors. In speedcubing, high turns per second (TPS), minimal pick-up and ending times, zero idle time [1], and One-Look Last Layer (1LLL) algorithms [2] are all critical to surpassing the current world record [3]. Among these, block building represents one of the most significant limiting factors [4, 5]. While numerous approaches exist for constructing a 2×2x2 or even a 2×2x3 block, determining an optimal block-building move sequence remains an open problem in speedcubing. This paper addresses the systematic construction of block building for the 3×3 Rubik's Cube. Here is the link of the AK-cubie-method software:
https://huggingface.co/spaces/AK51/AK_cubie_method
Video Link:
https://youtu.be/Z03pB7GI-Rw

## II. Background

This paper is the extension of the previous paper, "The Possibility of Solving a 3×3 Rubik's Cube Under 3 Seconds" [1]. To preclude redundancy, the fundamental principles of the Rubik's Cube are omitted in this paper. With the 3-second milestone having already been surpassed [3], the next objective is to eclipse the 2-second threshold. The current world record for the 3×3 Rubik's Cube stands at 2.76 seconds, achieved by 9-year-old Teodor Zajder at the GLS Big Cubes Gdańsk 2026 event in Poland on February 8, 2026 [6-8], with a move count of 29.

Within our software, selecting the Two-Phase (Kociemba) search [9] yields a move count of merely 16; however, the resultant move sequence is not easy for human to understand.

By contrast, the proposed **AK-cubie method** produces a move count of 25 while remaining human-readable.

Here is the scramble of the current world record (2.76 sec):
L B R2 B' R2 U2 F D R2 U R2 F2 D2 R U B L2.

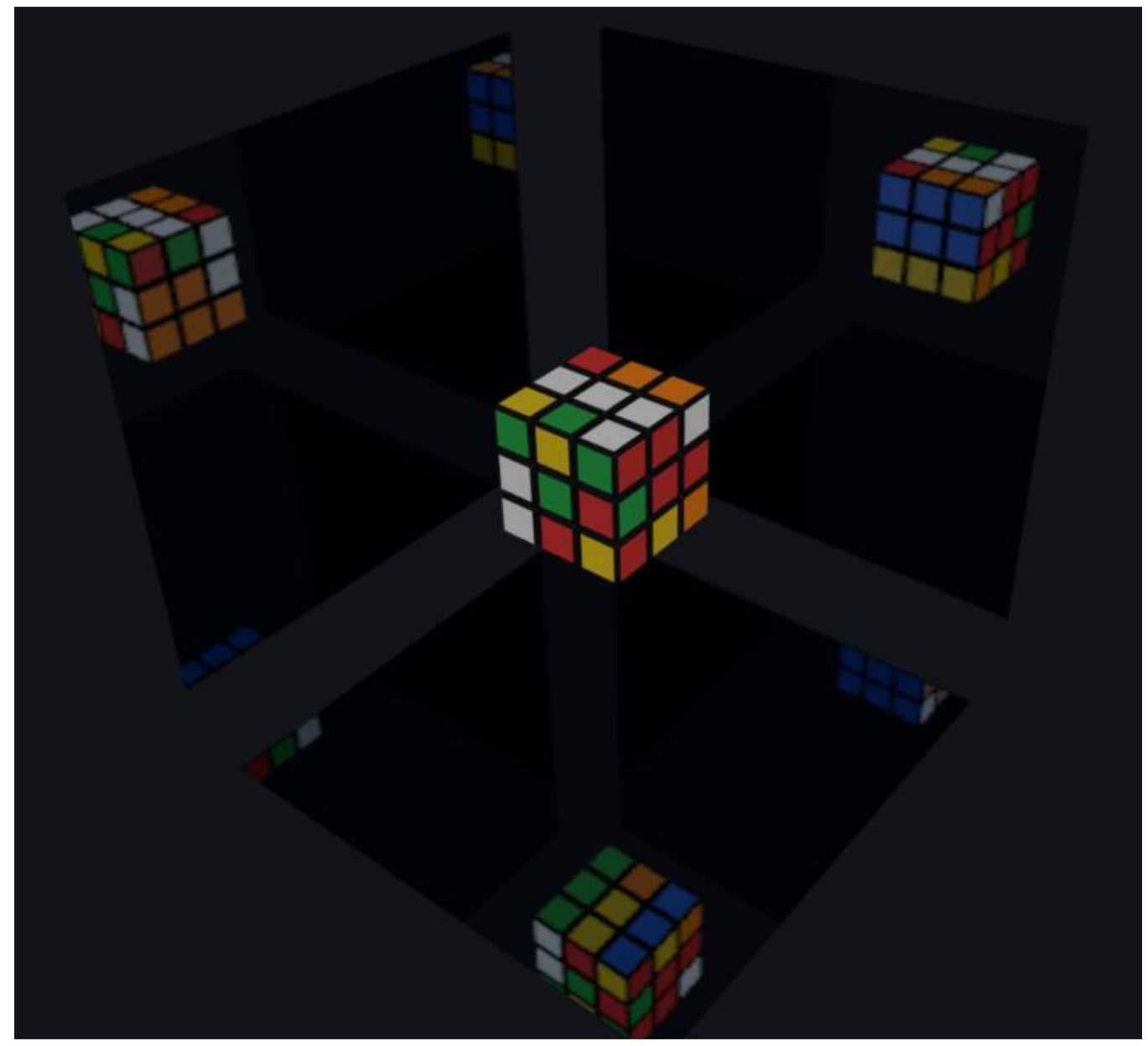

*Fig 2. world record Scramble cube*

Here is the solution from Teodor Zajder.
x' // inspection
r' U F U' r U' r' U2 r' U r // xxxcross
R U2' R2' U' R U R U2' R' // 4th pair
U' F' r U R' U' r' F R // ZBLL

For the Two-phase (Kociemba) search:
B2 R2 U2 R' B U' B2 R U B R' B2 L' R' D B, but it is hard for a human to understand this method.

For the new AK-cubie method, the total move count is 25 with 2 cube rotation.
l U2 y' R F D R2 y' R2 F' U2 F' U' F' U R U' R' U2 R' F R2 U R' U' F' U

The software breaks down the solution in several parts so that speedcubers can have a better understanding of the thinking process.

AK-cubie method — thinking process
1.Build the 2×2×2 block — green + white + red corner (6 moves)
l U2 y' R F D R2

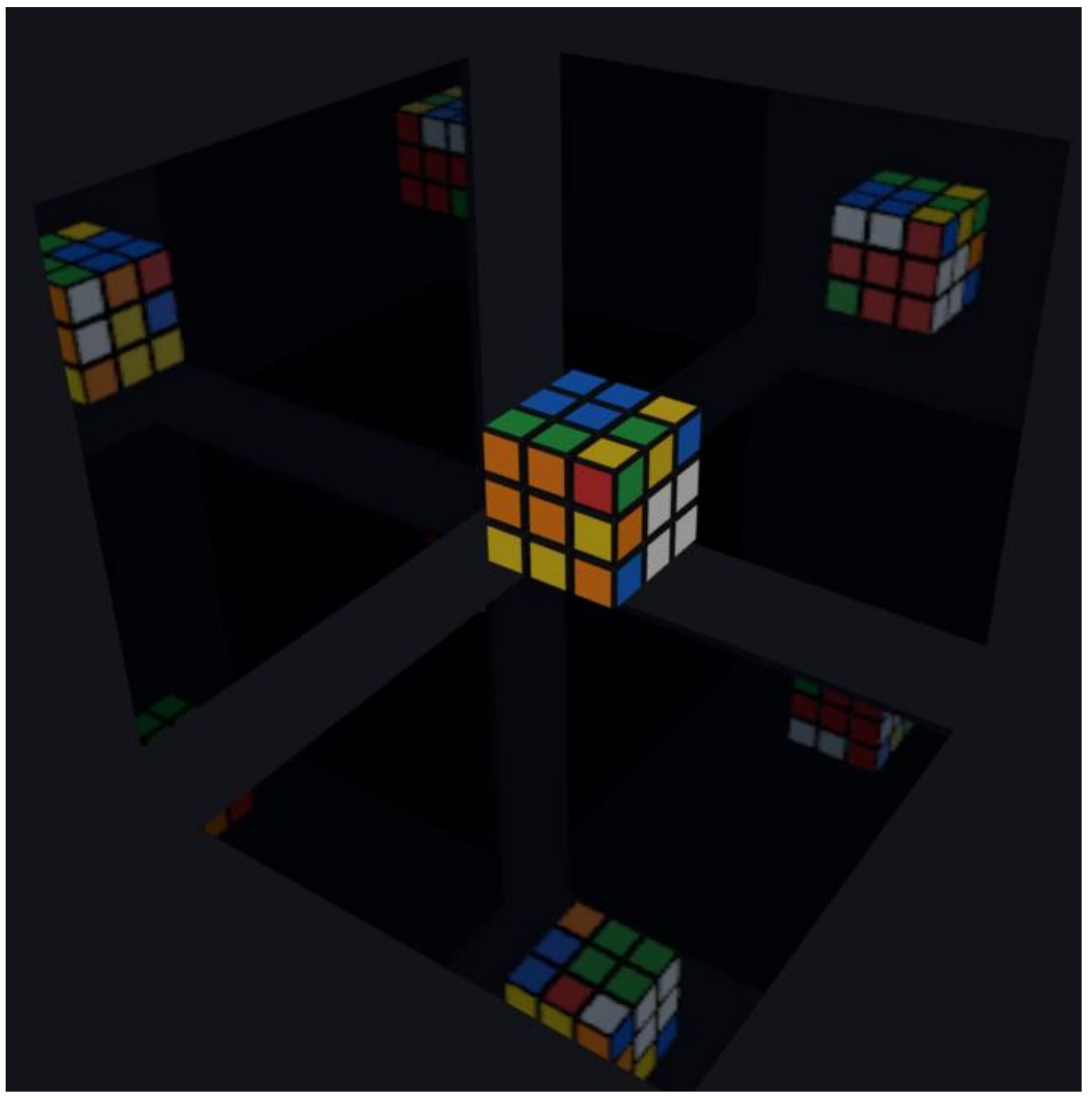

*Fig 3. 2x2x2 white green red block*

2.Extend to xxx-cross — adding the green + yellow + red corner (6 moves)
y' R2 F' U2 F' U' F'

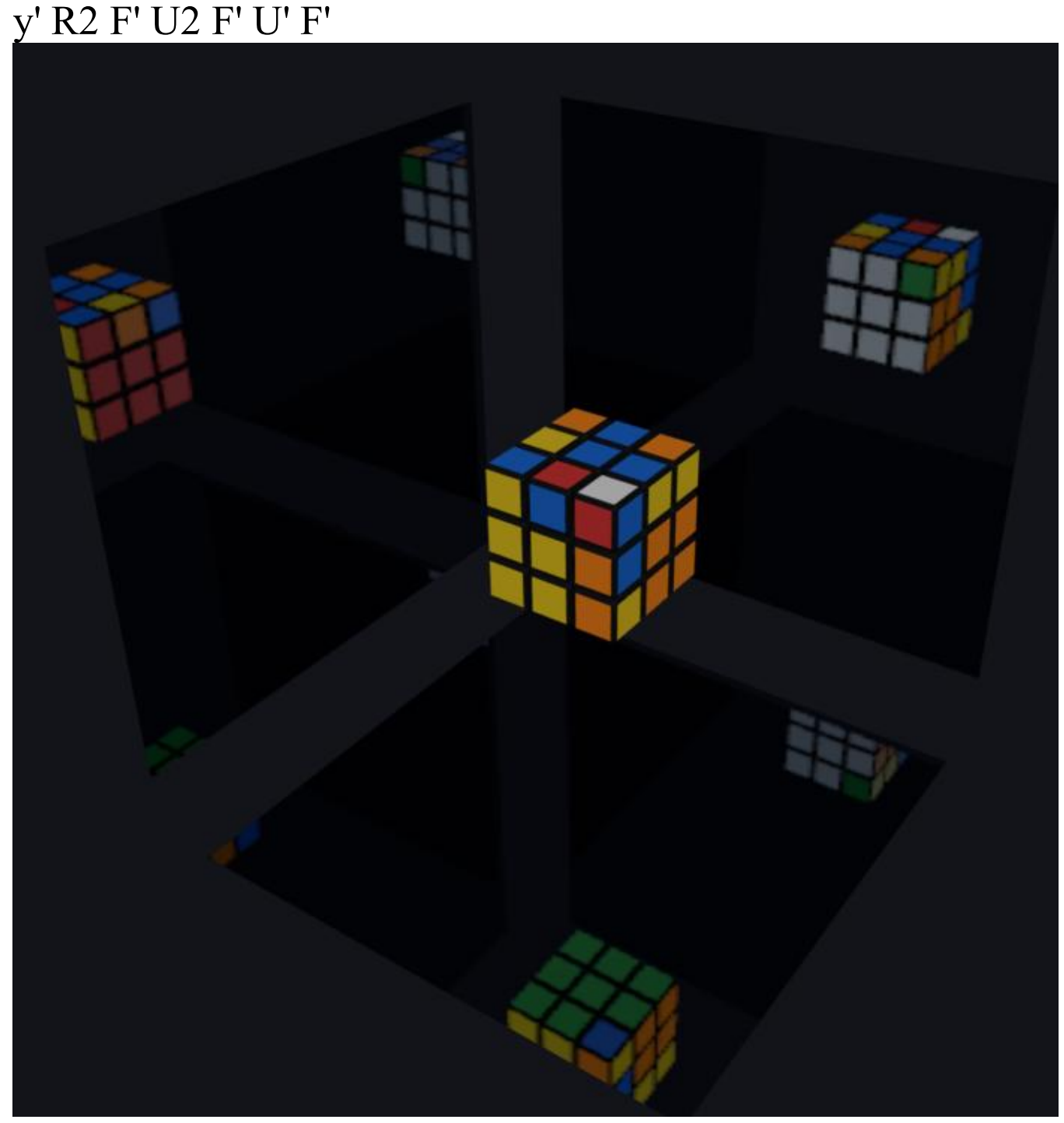

*Fig 4. xxx-cross with green bottom*

3.Complete the first two layers (F2L) (4 moves) – with move cancellation with 1LLL
U R

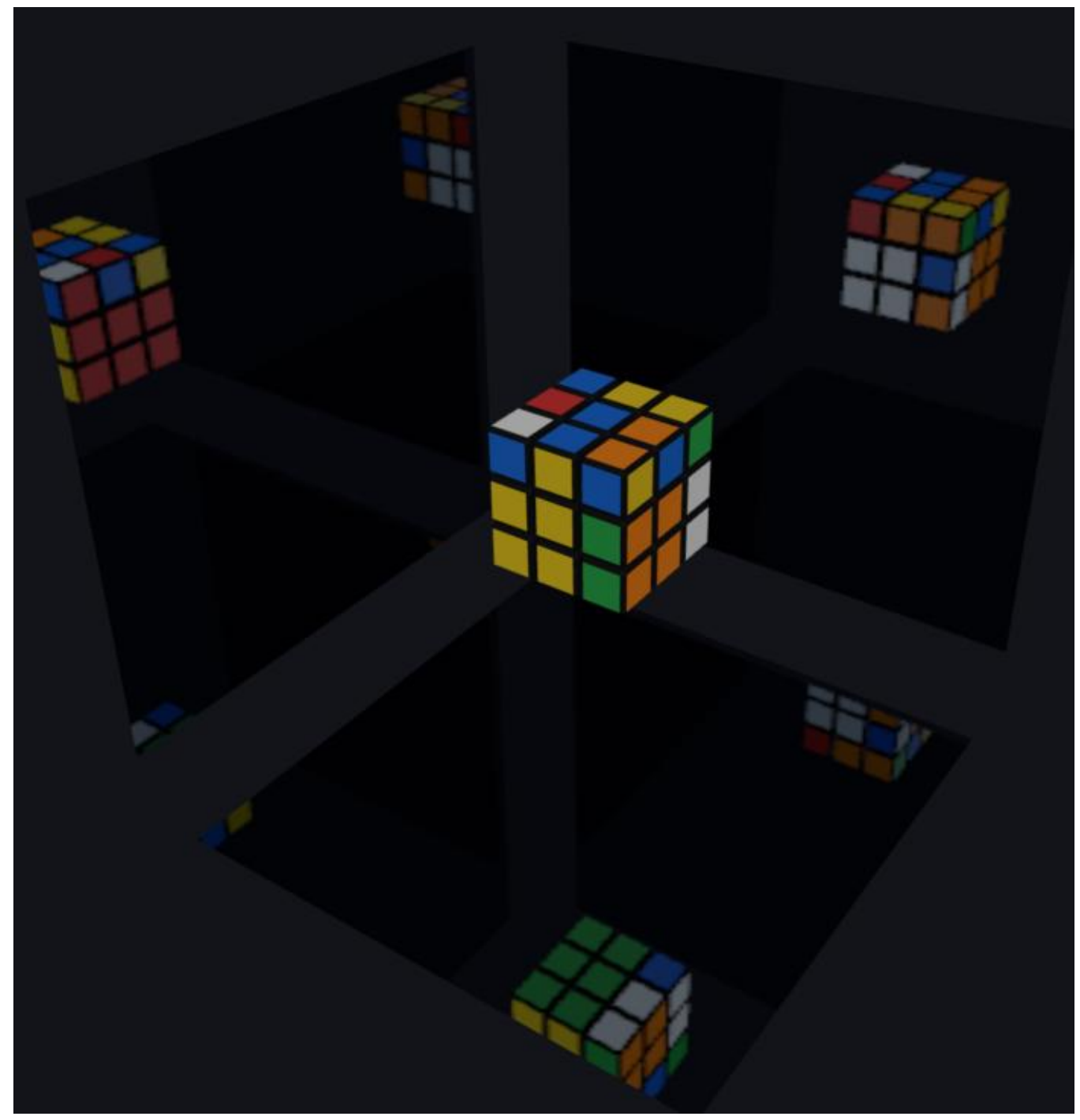

*Fig 5. Last F2L*

4.Last layer (1LLL) (12 moves)
U' R' U2 R' F R2 U R' U' F' U

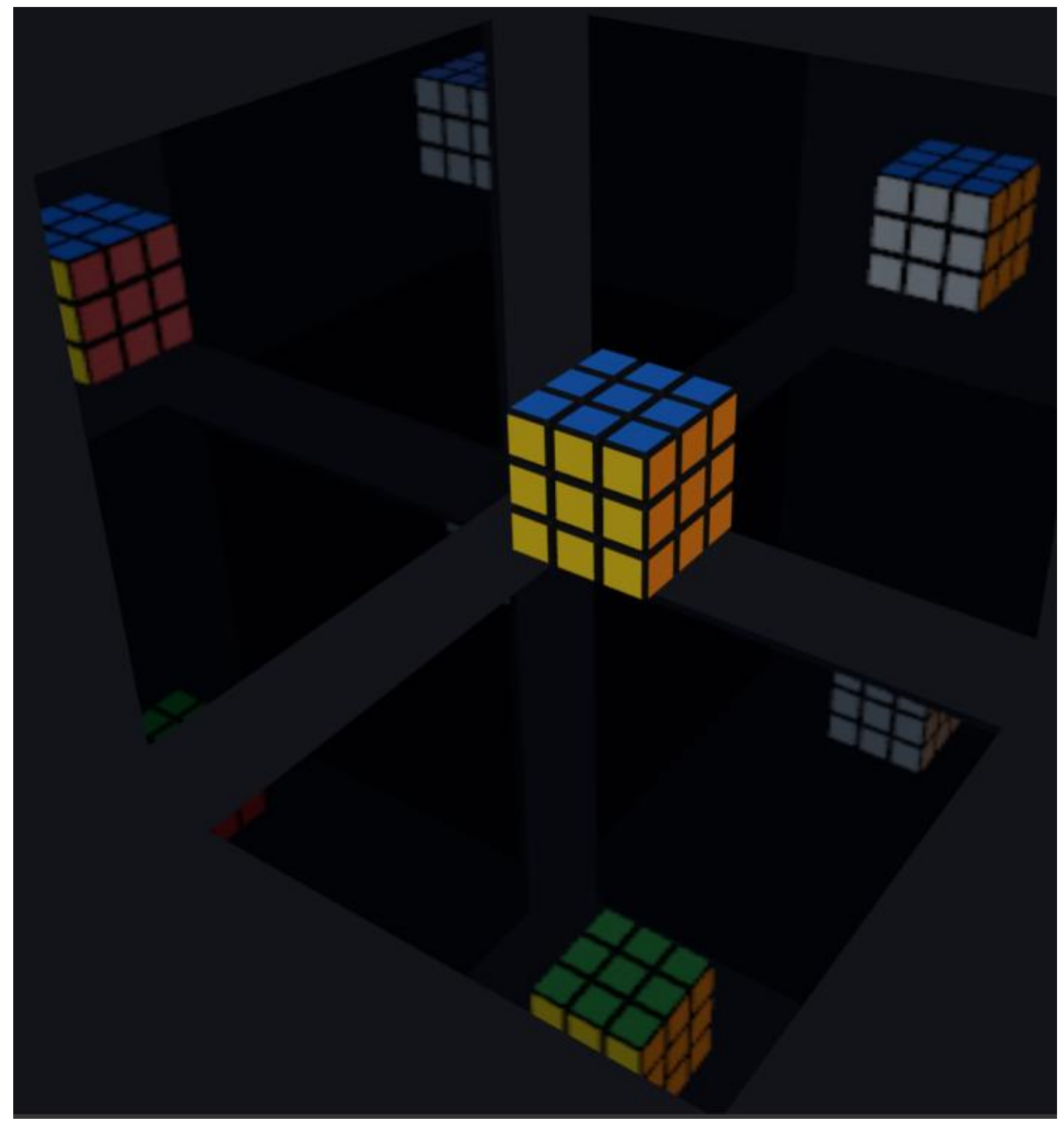

*Fig 6. Solved cube*

ONE-HAND (ROTATIONS + WIDE MOVES)

R x’ U2 y’ R F D R2 y’ R2 F’ U2 F’ U’ F’ U R U’ R’ U2 R’ F R2 U R’ U’ F’ U

TWO-HAND (WIDE + SLICE MOVES)

l U2 y’ R F D R2 y’ R2 F’ U2 F’ U’ F’ U R U’ R’ U2 R’ F R2 U R’ U’ F’ U

Same solve, re-notated for comfortable execution (see About). Ends fully solved, possibly in a rotated orientation.

AK-CUBIE METHOD — THINKING PROCESS

**1. Build the 2×2×2 block — green + white + red corner** (6 moves)

HTM R B2 U L F U2
OH R x’ U2 y’ R F D R2
2H l U2 y’ R F D R2

a 2×1 pair (corner + edge) forms after 6 moves

**2. Extend to a 2×2×3 block — adding the green + yellow + red corner** (6 moves)

HTM L2 D’ B2 D’ B’ D’
OH y’ R2 F’ U2 F’ U’ F’
2H y’ R2 F’ U2 F’ U’ F’

grow the solid block to fill two layers on one side

**3. Complete the first two layers (F2L)** (4 moves)

HTM B L B L’
OH U R
2H U R

add the remaining corner + edge pairs by block building

**4. Last layer (1LLL)** (12 moves)

HTM L B2 L’ B2 L’ D L2 B L’ B’ D’ B
OH U’ R’ U2 R’ F R2 U R’ U’ F’ U
2H U’ R’ U2 R’ F R2 U R’ U’ F’ U

orient and permute the whole last layer in one sequence

*Fig 7. AK-cubie Method for* 2.76 seconds’ record

The AK-cubie method generates three types of solutions: HTM, OH, and 2H. HTM denotes the solution with the fewest moves under the half-turn metric; OH denotes the one-handed solution; and 2H denotes the two-handed solution. The associated finger tricks are detailed in the following section. In addition to the current world record, the two preceding world records have also been tested using the AK-cubie method.

For the world record of 3.05 seconds by Xuanyi Geng on 2025 April 13 [10, 11].
Scramble: R2 D2 R2 D2 U' F2 D L2 B2 R' D R B2 U2 L R2 D U2 F R2

Xuanyi Geng’s solution
x2 // inspection
l2 F L' U' R // cross
U R' U2' R // 1st pair
U L U' L' // 2nd pair

U L' U' L // 3rd pair
R U2' R' U' R U R' // 4th pair
F R' F' r U R U' r' U // ZBLL

Below is the AK-cubie method

ONE-HAND (ROTATIONS + WIDE MOVES)
y' R' y' R' D z U' R' D F2 R U R' y' F2 U2 R2 F' z' R2 D' R D2 R' u' U2 F2 R y R2 F2 U2 R2 y D2 R'

TWO-HAND (WIDE + SLICE MOVES)
F' L' D R' U' L y R2 U F U' z' R2 F2 U2 R' y R2 D' R D2 R' D' U2 L2 F R2 F2 U2 R2 D2 y R'

Same solve, re-notated for comfortable execution (see About). Ends fully solved, possibly in a rotated orientation.

AK-CUBIE METHOD — THINKING PROCESS

**1. Build the 2×2×2 block — white + green + orange corner** (6 moves)
HTM F' L' D R' U' L
OH y' R' y' R' D z U' R' D
2H F' L' D R' U' L
a 2×1 pair (corner + edge) forms after 6 moves

**2. Extend to a 2×2×3 block — adding the white + blue + orange corner** (4 moves)
HTM B2 U R U'
OH F2 R U R'
2H y R2 U F U'
grow the solid block to fill two layers on one side

**3. Complete the first two layers (F2L)** (2 moves)
HTM D2 R2
OH y' F2 U2
2H z' R2 F2
add the remaining corner + edge pairs by block building

**4. Last layer (1LLL)** (17 moves)
HTM B2 D' L2 F' L F2 L' F' B2 R2 D L2 D2 B2 L2 F2 U'
OH R2 F' z' R2 D' R D2 R' u' U2 F2 R y R2 F2 U2 R2 y D2 R'
2H U2 R' y R2 D' R D2 R' D' U2 L2 F R2 F2 U2 R2 D2 y R'
orient and permute the whole last layer in one sequence

*Fig 8. AK-cubie Method for* 3.05 seconds' record

For the world record of 3.08 seconds by Yiheng Wang on 2025 February 16 [12, 13].

Scramble: U2 R' D2 R B2 D2 B2 R2 B F U F R2 B2 R F' L2 F2 L

Yiheng Wang's solution

z' y // inspection
U' D' r R' D U' R' U' D // cross
R U R' // 1st pair
L U L' U' L U L2' // 2nd pair
U' L // 3rd pair
U2 R' U R U' R' U R // 4th pair
U' R' U R' U' R3 U' R' U R U R' U' U R' U' // EPLL

Below is the AK-cubie method

ONE-HAND (ROTATIONS + WIDE MOVES)
F2 D2 z U' D' R' F R2 U2 x' U D R' D' U2 R' U2 R2 U2 x U D' R2 U' D R2 y' R' F U2 F' R y R y R2 F R2

TWO-HAND (WIDE + SLICE MOVES)
F2 D2 L' R' U' F U2 L2 y R L U' L' R2 U' R2 U2 R2 y' L R' U2 L' R U2 F' D L2 D' F U y R2 U R2

Same solve, re-notated for comfortable execution (see About). Ends fully solved, possibly in a rotated orientation.

AK-CUBIE METHOD — THINKING PROCESS

**1. Build the 2×2×2 block — green + red + yellow corner** (6 moves)
HTM F2 D2 L' R' U' F
OH F2 D2 z U' D' R' F
2H F2 D2 L' R' U' F
a 2×1 pair (corner + edge) forms after 6 moves

**2. Extend to a 2×2×3 block — adding the green + orange + yellow corner** (6 moves)
HTM U2 L2 B F U' F'
OH R2 U2 x' U D R' D'
2H U2 L2 y R L U' L'
grow the solid block to fill two layers on one side

**3. Complete the first two layers (F2L)** (10 moves)
HTM B2 U' B2 U2 B2 L R' U2 L' R
OH U2 R' U2 R2 U2 x U D' R2 U' D
2H R2 U' R2 U2 R2 y' L R' U2 L' R
add the remaining corner + edge pairs by block building

**4. Last layer (1LLL)** (10 moves)
HTM U2 F' D L2 D' F U B2 U B2
OH R2 y' R' F U2 F' R y R y R2 F R2
2H U2 F' D L2 D' F U y R2 U R2
orient and permute the whole last layer in one sequence

*Fig 9. AK-cubie Method for* 3.08 seconds' record

## III. SCOPE OF WORK

Rather than analyzing the entire 3×3 Rubik's Cube, the AK-cubie method decomposes the cube into 6 center cubies, 12

edge cubies, and 8 corner cubies. The center cubies can be disregarded, as no wide moves or cube rotations occur within the core algorithm. Ergonomic solutions for one-handed and two-handed solves are generated after an optimal path has been found.

The core algorithm employs only the following moves:
R, L, U, D, F, B, R′, L′, U′, D′, F′, B′, R2, L2, U2, D2, F2, B2

Regarding cube orientation within the core algorithm, the white center cubie is always positioned at the bottom, and the blue center cubie faces the user. To facilitate visualization of the scramble algorithm, the software's user interface rotates the cube so that white appears on top and green faces the front. A detailed explanation of the core algorithm follows.

**Edge position.** The 12 edge positions are indexed as follows: UF = 0, UR = 1, UB = 2, UL = 3, FR = 4, BR = 5, BL = 6, FL = 7, DF = 8, DR = 9, DB = 10, and DL = 11, where U, R, L, D, F, and B denote the up, right, left, down, front, and back faces, respectively. For example, on a solved cube, the yellow-blue edge occupies UF, and the blue-orange edge occupies FL.

**Edge orientation.** An edge cubie may assume one of two orientation states: good or bad. A good edge can be placed into its correct position and orientation using only R, L, U, and D. A bad edge requires F or B to be converted into a good edge. In the algorithm, a value of 0 denotes a good edge, while 1 denotes a bad edge.

**Corner position.** The eight corner positions are indexed as follows: UFR = 0, UBR = 1, UBL = 2, UFL = 3, DFR = 4, DBR = 5, DBL = 6, and DFL = 7, where U, R, L, D, F, and B denote the up, right, left, down, front, and back faces, respectively. For example, on a solved cube, the yellow-blue-red corner occupies UFR, and the white-blue-orange corner occupies DFL.

**Corner orientation.** A corner cubie may assume one of three orientation states, determined by the position of its white or yellow sticker. If the white or yellow sticker faces up or down, the corner is classified as good. If it faces left or right, the corner is classified as side. If it faces the front or back, the corner is classified as face. In the AK-cubie algorithm, 0 denotes as a good corner, 1 as a side corner, and 2 as a face corner.

Inside the AK-cubie algorithm, the edge cubies and corner cubies are treated separately. There are four arrays: position of edge cubies, orientation of edge cubies, position of corner cubies, orientation of corner cubies. For the solved cube, here is the **AK-cubie skeleton**:
(0,1,2,3,4,5,6,7,8,9,10,11) (0,0,0,0,0,0,0,0,0,0,0,0)
(0,1,2,3,4,5,6,7) (0,0,0,0,0,0,0,0)
They are (edge position)(edge orientation)(corner position)(corner orientation)

e.g. For a cube that needs a U move to solve it, the AK-cubie skeleton is
(3,0,1,2,4,5,6,7,8,9,10,11)(0,0,0,0,0,0,0,0,0,0,0,0)
(3,0,1,2,4,5,6,7) (0,0,0,0,0,0,0,0)

e.g. For a cube that needs a R move to solve it, the AK-cubie skeleton is
(0,5,2,3,1,9,6,7,8,4,10,11)(0,0,0,0,0,0,0,0,0,0,0,0)
(1,5,2,3,0,4,6,7) (2,2,0,0,2,2,0,0)

e.g. For a cube that needs a F move to solve it, the AK-cubie skeleton is
(4,1,2,3,8,5,6,0,7,9,10,11) (1,0,0,0,1,0,0,1,1,0,0,0)
(4,1,2,0,7,5,6,3) (1,0,0,1,1,0,0,1)

The algorithm solves one cubie at a time, combining cubies incrementally until eight edge cubies and four corner cubies are solved, thereby completing the first two layers. At most three moves are required to place any edge cubie into its correct position and orientation, and likewise at most three moves are required for any corner cubie. The symbol "x" denotes a "don't care" condition. The one-cubie solutions are illustrated below.

e.g. For Yellow orange edge, if it is position is at UF and orientation is good, then a U move is enough.
(3,x,x,x,x,x,x,x,x,x,x,x)(0,x,x,x,x,x,x,x,x,x,x,x)
(x,x,x,x,x,x,x,x) (x,x,x,x,x,x,x,x)
Move list
U

e.g.
For Yellow blue edge, position is correct, orientation is bad
(0,x,x,x,x,x,x,x,x,x,x,x)(1,x,x,x,x,x,x,x,x,x,x,x)
(x,x,x,x,x,x,x,x) (x,x,x,x,x,x,x,x)
Move list
F,R,U
F',L',U'
U',R',F'
U,L,F

e.g.
For Yellow blue red corner, position is correct, orientation is face
(x,x,x,x,x,x,x,x,x,x,x,x)(x,x,x,x,x,x,x,x,x,x,x,x)
(0,x,x,x,x,x,x,x) (2,x,x,x,x,x,x,x)
Move list
R,U
R',F2,U'
F',L',U2
F,R
U2,L,U'
U,F

First, list out all move-sequence list for each edge cubie and each corner cubie to the correct position and orientation from any position and orientation. The maximum number of moves is 3.

Edge cubie from any position and orientation:
(11 x 2 + 1) for one edge cubie, 11 wrong positions, 2 orientations, and 1 right position but wrong orientation.
As there are 12 edge cubies, the move list is the same except the orientation of the whole cube is different.
12 (11 x 2 + 1) = 276 move groups for all 12 edge cubies

Corner cubie from any position and orientation:
(7 x 3 + 2) for one corner cubie, 7 wrong positions, 3 orientations, and 2 right positions but wrong orientation.
As there are 8 corner cubies, the move list is the same except the orientation of the whole cube is different.
8 (7 x 3 + 2) = 184 move groups for all 8 corner cubies

Before adding a new cubie to the move sequence, we have to look into the existing move sequence, if we pick a move sequence of " R,U ", we don't care about L,D,F,B move if that cubie is not affected by these moves.

e.g. if we combine yellow orange edge at UF with correct orientation, and yellow blue orange corner cubie at UFR with correct orientation

yellow orange edge at UF with correct orientation
(3,x,x,x,x,x,x,x,x,x,x,x)(0,x,x,x,x,x,x,x,x,x,x,x)
(x,x,x,x,x,x,x,x) (x,x,x,x,x,x,x,x)
Move List:
U

yellow blue orange corner at UFR with correct orientation
(x,x,x,x,x,x,x,x,x,x,x,x)(x,x,x,x,x,x,x,x,x,x,x,x)
(3,x,x,x,x,x,x,x) (0,x,x,x,x,x,x,x)
Move List:
U

Then, the combined move is just U.

Another example
For Yellow blue edge, position is correct, orientation is bad
(0,x,x,x,x,x,x,x,x,x,x,x)(1,x,x,x,x,x,x,x,x,x,x,x)
(x,x,x,x,x,x,x,x) (x,x,x,x,x,x,x,x)
Move list
F,R,U
F',L',U'
U',R',F'
U,L,F

For Yellow blue red corner, position is correct, orientation is face
(x,x,x,x,x,x,x,x,x,x,x,x)(x,x,x,x,x,x,x,x,x,x,x,x)
(0,x,x,x,x,x,x,x) (2,x,x,x,x,x,x,x)
Move list
R,U
R',F2,U'
F',L',U2
F,R
U2,L,U'
U,F

The thinking process is this, if we pick F',L',U' from the edge cubie, we can add don't care "x" into the moves sequence as long as the additional move does not affect the previous cubies we care about. e.g. F',x,L',U',x and then, try different combinations for the top 50 shortest move sequences. One of the combined move sequence can be F',U',L',U',R'. This trimming value 50 can limit the branching of the tree in each cubie adding. If the RAM or memory usage is getting high, the program will halt. By combining one cubie at a time, it is possible to get the lowest count moves for block building.

**Selecting the first cubie**
The Petrus method [14] is based on block building. The newly proposed AK-cubie method adopts a similar concept, beginning with the construction of a 2×1 block and progressively extending it into 2×2 and 2×3 blocks. As the AK-cubie method can be executed and optimized computationally, it can evaluate all 24 possible 2×1 blocks corresponding to different color combinations, i.e., 4 F2L pairs × 6 bottom colors = 24.

For each solution, the AK-cubie method additionally employs NISS (inverse scramble) and insertion techniques to further reduce the move count. During the transition from block building to 1LLL, move cancellations may occur, yielding additional move reductions. Achieving such move reduction poses a significant challenge, as most speedcubers are accustomed to commencing the 1LLL from a consistent initial setup due to muscle memory.

Once a solution is identified, the move sequence is subsequently modified to produce finger-friendly sequences for both two-handed and one-handed solving. Since the lowest move count does not always correspond to a finger-friendly sequence, the AK-cubie software presents the top 10 solutions with the lowest move counts, allowing the user to make the final selection.

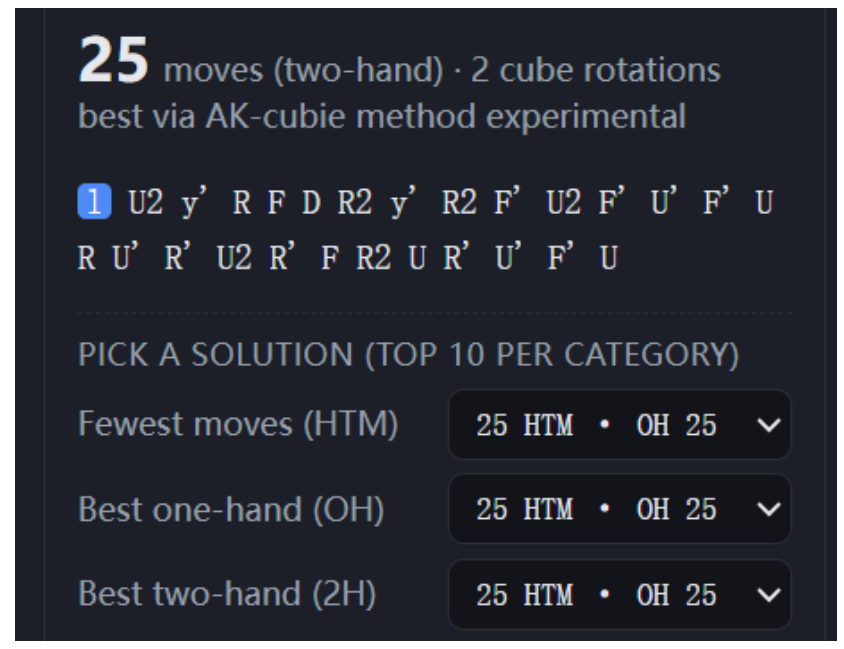

*Fig 10. Top 10 records per category*

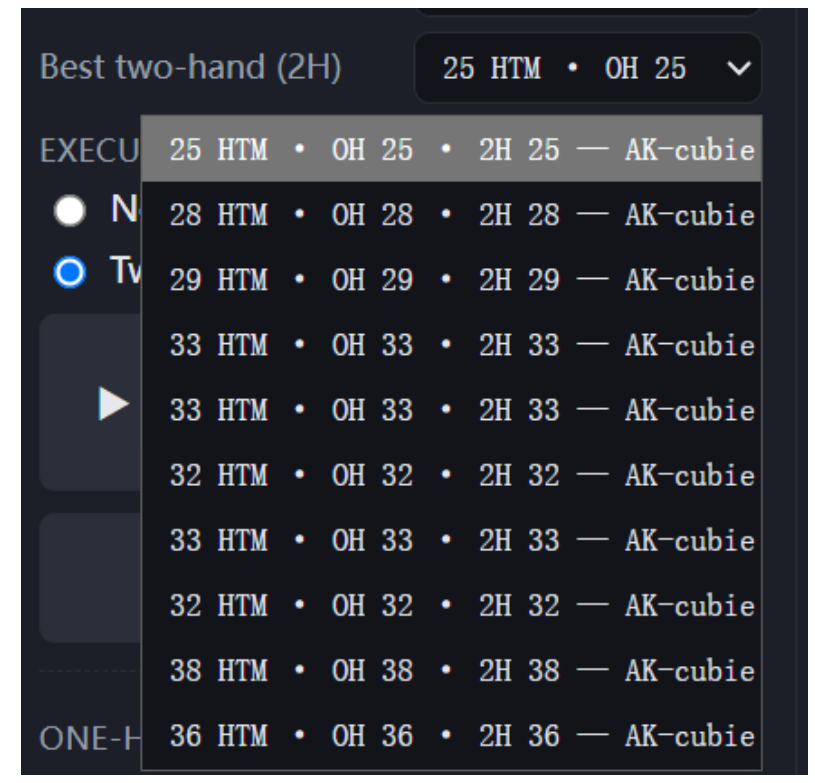

*Fig 11. Top 10 two-hands solutions*

The entire solving program completes within 300 seconds and yields optimal solutions, with this default time limit being adjustable. Users may select either the Two-phase (Kociemba) search or the AK-cubie method. The computation time is hardware-dependent; the present benchmark was conducted on a laptop equipped with an RTX 4070 GPU. The software provides numerous settings that users can adjust to suit their testing environment.

## IV. Test Result

Example 1
Rubik's Cube Solve
Saved: Wed Sep 09 2026 22:49:53 GMT+0800 (香港标准时间)
Strategy: AK-cubie method experimental
All solves start with WHITE on top and GREEN in front.

Scramble:
F R' F U D B2 D L' R2 B' D2 B' R' L D2 B' U R2 F L' R' B' L2 F D2

Normal (HTM) — 25 moves:
B2 U L' F2 B2 D' B U' B' U R' U2 B' R F2 R' B2 U2 F' U' R U' R' B' R

One-hand (rotations + wide moves) — 25 moves (+5 cube rotations):
y R2 U y R' x D2 U2 F' y U R' U' R y R' F2 U' R D2 R' U2 F2 D' F' R F' R' U' R

Two-hand (wide + slice moves) — 25 moves (+3 cube rotations):
y R2 U y R' y' L2 R2 D' R U' R' U F' U2 R' F L2 F' R2 U2 L' U' F U' F' R' F

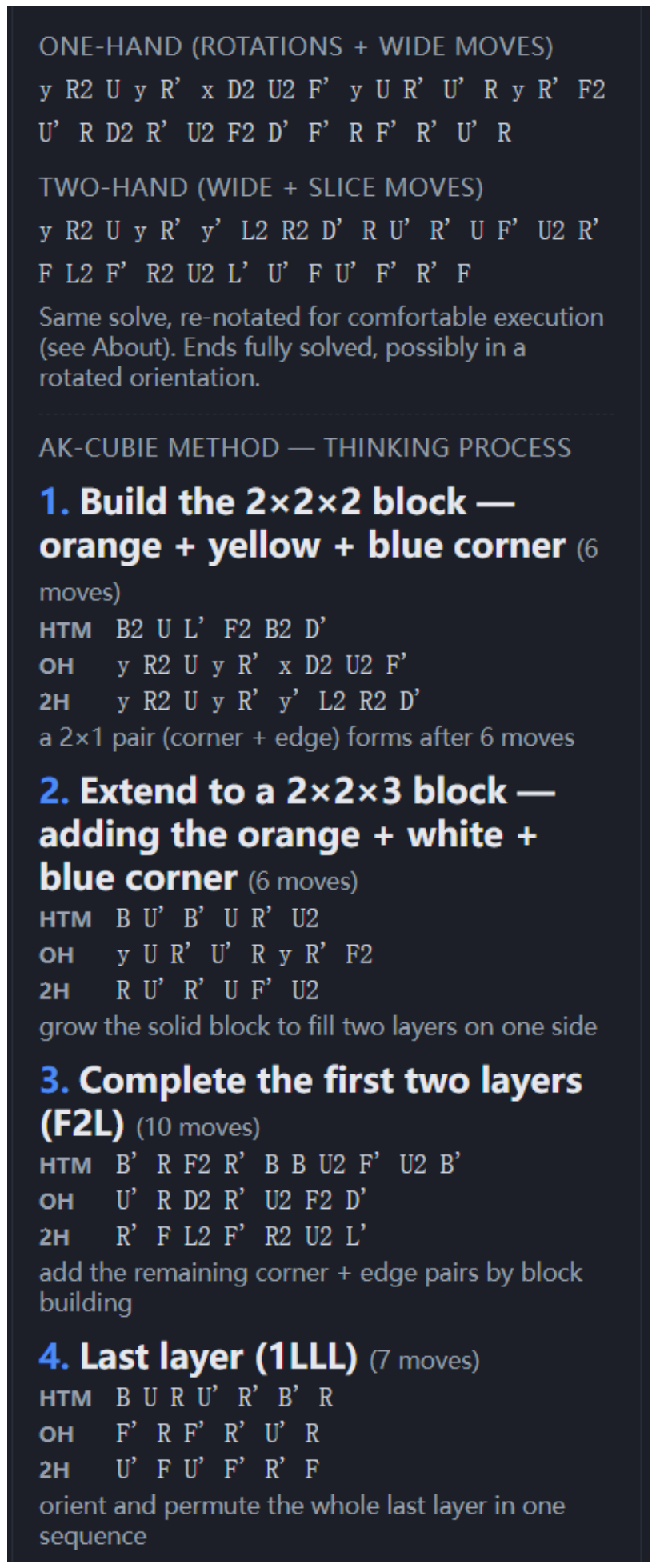

*Fig 12. AK-cubie Method*

Example 2
Rubik's Cube Solve
Saved: Wed Sep 09 2026 23:01:30 GMT+0800 (香港标准时间)
Strategy: AK-cubie method experimental
All solves start with WHITE on top and GREEN in front.

Scramble:
D2 R2 F' D2 L D2 B' F2 U' B' D' U' B L U' R L2 D2 R' D F2 D' U R' D

Normal (HTM) — 36 moves:
F R B' D B2 F2 L F' D2 L2 D' L2 F' L F L2 F2 D F D' U L U' D2 U' F D2 F' D2 F U' F2 U2 B2 D2 B2

One-hand (rotations + wide moves) — 36 moves (+5 cube rotations):
F R y R' D R2 y' F2 y' F R' D2 F2 D' F2 R' F R F2 R2 D R D' U F U' D2 U' R D2 R' D2 R U' R2 y U2 y R2 D2 R2

Two-hand (wide + slice moves) — 36 moves (+1 cube rotation):
F R y' L' D L2 R2 F R' D2 F2 D' F2 R' F R F2 R2 D R D' U F U' D2 U' R D2 R' D2 R U' R2 U2 L2 D2 L2

ONE-HAND (ROTATIONS + WIDE MOVES)
F R y R’ D R2 y’ F2 y’ F R’ D2 F2 D’ F2 R’ F R F2 R2 D R D’ U F U’ D2 U’ R D2 R’ D2 R U’ R2 y U2 y R2 D2 R2

TWO-HAND (WIDE + SLICE MOVES)
F R y’ L’ D L2 R2 F R’ D2 F2 D’ F2 R’ F R F2 R2 D R D’ U F U’ D2 U’ R D2 R’ D2 R U’ R2 U2 L2 D2 L2

Same solve, re-notated for comfortable execution (see About). Ends fully solved, possibly in a rotated orientation.

AK-CUBIE METHOD — THINKING PROCESS

**1. Build the 2×2×2 block — red + white + blue corner** (5 moves)
HTM F R B’ D B2
OH F R y R’ D R2
2H F R y’ L’ D L2
a 2×1 pair (corner + edge) forms after 5 moves

**2. Extend to a 2×2×3 block — adding the red + yellow + blue corner** (6 moves)
HTM F2 L F’ D2 L2 D’
OH y’ F2 y’ F R’ D2 F2 D’
2H R2 F R’ D2 F2 D’
grow the solid block to fill two layers on one side

**3. Complete the first two layers (F2L)** (12 moves)
HTM L2 F’ L F L2 F2 D F D’ U L U’
OH F2 R’ F R F2 R2 D R D’ U F U’
2H F2 R’ F R F2 R2 D R D’ U F U’
add the remaining corner + edge pairs by block building

**4. Last layer (1LLL)** (13 moves)
HTM D2 U’ F D2 F’ D2 F U’ F2 U2 B2 D2 B2
OH D2 U’ R D2 R’ D2 R U’ R2 y U2 y R2 D2 R2
2H D2 U’ R D2 R’ D2 R U’ R2 U2 L2 D2 L2
orient and permute the whole last layer in one sequence

*Fig 13. AK-cubie Method*

Example 3
Rubik's Cube Solve
Saved: Wed Sep 09 2026 23:27:13 GMT+0800 (香港标准时间)
Strategy: AK-cubie method experimental
All solves start with WHITE on top and GREEN in front.

Scramble:
F U2 B2 L2 D' B D' F' B2 D2 R' F' L D' U B' U F2 B' L B R2 F2 L' U

Normal (HTM) — 30 moves:
U2 R B R' B2 U' L2 F L D2 F2 L' B' L' F L B D' F2 D L2 D2 R' D' R D2 L' D L' F'

One-hand (rotations + wide moves) — 30 moves (+5 cube rotations):
U2 R x' U R' U2 y' R' y' R2 D R F2 D2 R' U' R' D R z R y' R' F2 R D2 R2 U' R' U R2 D' R D' F'

Two-hand (wide + slice moves) — 30 moves (+2 cube rotations):
U2 l U R' U2 y' R' y' R2 D R F2 D2 R' U' R' D R U F' D2 F R2 F2 L' F' L F2 R' F R' D'

ONE-HAND (ROTATIONS + WIDE MOVES)
U2 R x’ U R’ U2 y’ R’ y’ R2 D R F2 D2 R’ U’ R’ D R z R y’ R’ F2 R D2 R2 U’ R’ U R2 D’ R D’ F’

TWO-HAND (WIDE + SLICE MOVES)
U2 l U R’ U2 y’ R’ y’ R2 D R F2 D2 R’ U’ R’ D R U F’ D2 F R2 F2 L’ F’ L F2 R’ F R’ D’

Same solve, re-notated for comfortable execution (see About). Ends fully solved, possibly in a rotated orientation.

AK-CUBIE METHOD — THINKING PROCESS

**1. Build the 2×2×2 block — blue + red + white corner** (6 moves)
HTM U2 R B R’ B2 U’
OH U2 R x’ U R’ U2 y’ R’
2H U2 l U R’ U2 y’ R’
a 2×1 pair (corner + edge) forms after 6 moves

**2. Extend to a 2×2×3 block — adding the blue + orange + white corner** (6 moves)
HTM L2 F L D2 F2 L’
OH y’ R2 D R F2 D2 R’
2H y’ R2 D R F2 D2 R’
grow the solid block to fill two layers on one side

**3. Complete the first two layers (F2L)** (8 moves)
HTM B’ L’ F L B D’ F2 D
OH U’ R’ D R z R y’ R’ F2 R
2H U’ R’ D R U F’ D2 F
add the remaining corner + edge pairs by block building

**4. Last layer (1LLL)** (10 moves)
HTM L2 D2 R’ D’ R D2 L’ D L’ F’
OH D2 R2 U’ R’ U R2 D’ R D’ F’
2H R2 F2 L’ F’ L F2 R’ F R’ D’
orient and permute the whole last layer in one sequence

*Fig 14. AK-cubie Method*

Example 4
Rubik's Cube Solve
Saved: Wed Sep 09 2026 23:40:48 GMT+0800 (香港标准时间)
Strategy: AK-cubie method experimental
All solves start with WHITE on top and GREEN in front.

Scramble:
L' B2 U2 D L' D' B R L F' D2 U B2 D F' D2 F' B D2 F B U B D R

Normal (HTM) — 32 moves:
B R' B' U R2 U' D L' D2 R' D R D F2 D2 L' D2 L F L' F D' F' D2 B D B2 D F D' B D

One-hand (rotations + wide moves) — 32 moves (+4 cube rotations):
y R F' R' y' U R2 U' z' R D' R2 U' R U R x U2 R2 F' R2 F U F' U R' U' R2 D R D2 R U R' D R

Two-hand (wide + slice moves) — 32 moves (+3 cube rotations):
y R F' R' d R2 U' z' R D' R2 U' R U R x U2 R2 F' R2 F U F' U R' U' R2 D R D2 R U R' D R

ONE-HAND (ROTATIONS + WIDE MOVES)
y R F’ R’ y’ U R2 U’ z’ R D’ R2 U’ R U R x U2 R2 F’ R2 F U F’ U R’ U’ R2 D R D2 R U R’ D R

TWO-HAND (WIDE + SLICE MOVES)
y R F’ R’ d R2 U’ z’ R D’ R2 U’ R U R x U2 R2 F’ R2 F U F’ U R’ U’ R2 D R D2 R U R’ D R

Same solve, re-notated for comfortable execution (see About). Ends fully solved, possibly in a rotated orientation.

AK-CUBIE METHOD — THINKING PROCESS

**1. Build the 2×2×2 block — red + white + blue corner** (6 moves)
HTM B R’ B’ U R2 U’
OH y R F’ R’ y’ U R2 U’
2H y R F’ R’ d R2 U’
a 2×1 pair (corner + edge) forms after 6 moves

**2. Extend to a 2×2×3 block — adding the red + yellow + blue corner** (6 moves)
HTM D L’ D2 R’ D R
OH z’ R D’ R2 U’ R U
2H z’ R D’ R2 U’ R U
grow the solid block to fill two layers on one side

**3. Complete the first two layers (F2L)** (11 moves)
HTM D F2 D2 L’ D D L F L’ F D’
OH R x U2 R2 F’ R2 F U F’ U R’
2H R x U2 R2 F’ R2 F U F’ U R’
add the remaining corner + edge pairs by block building

**4. Last layer (1LLL)** (10 moves)
HTM F’ D2 B D B2 D F D’ B D
OH U’ R2 D R D2 R U R’ D R
2H U’ R2 D R D2 R U R’ D R
orient and permute the whole last layer in one sequence

*Fig 15. AK-cubie Method*

Example 5
Rubik's Cube Solve
Saved: Wed Sep 09 2026 23:53:52 GMT+0800 (香港标准时间)
Strategy: AK-cubie method experimental
All solves start with WHITE on top and GREEN in front.

Scramble:
U' D2 F' U R B F R' D2 R2 L' D L' B' L2 R2 U R2 F R L' U' F' B U

Normal (HTM) — 32 moves:
F' L U' B D R2 U' B R U B' R' B R U B R' B2 R U R U2 L' U D' B' D B U' L U B'

One-hand (rotations + wide moves) — 32 moves (+4 cube rotations):
y' R' y' R y' U' R D F2 U' R F U R' F' R F U R F' R2 F U F U2 y R' U D' F' D F U' R U F'

Two-hand (wide + slice moves) — 33 moves:
F' L d' R D F2 U' R F U R' F' R F U R F' R2 F U F U' d' R' U D' F' D F U' R U F'

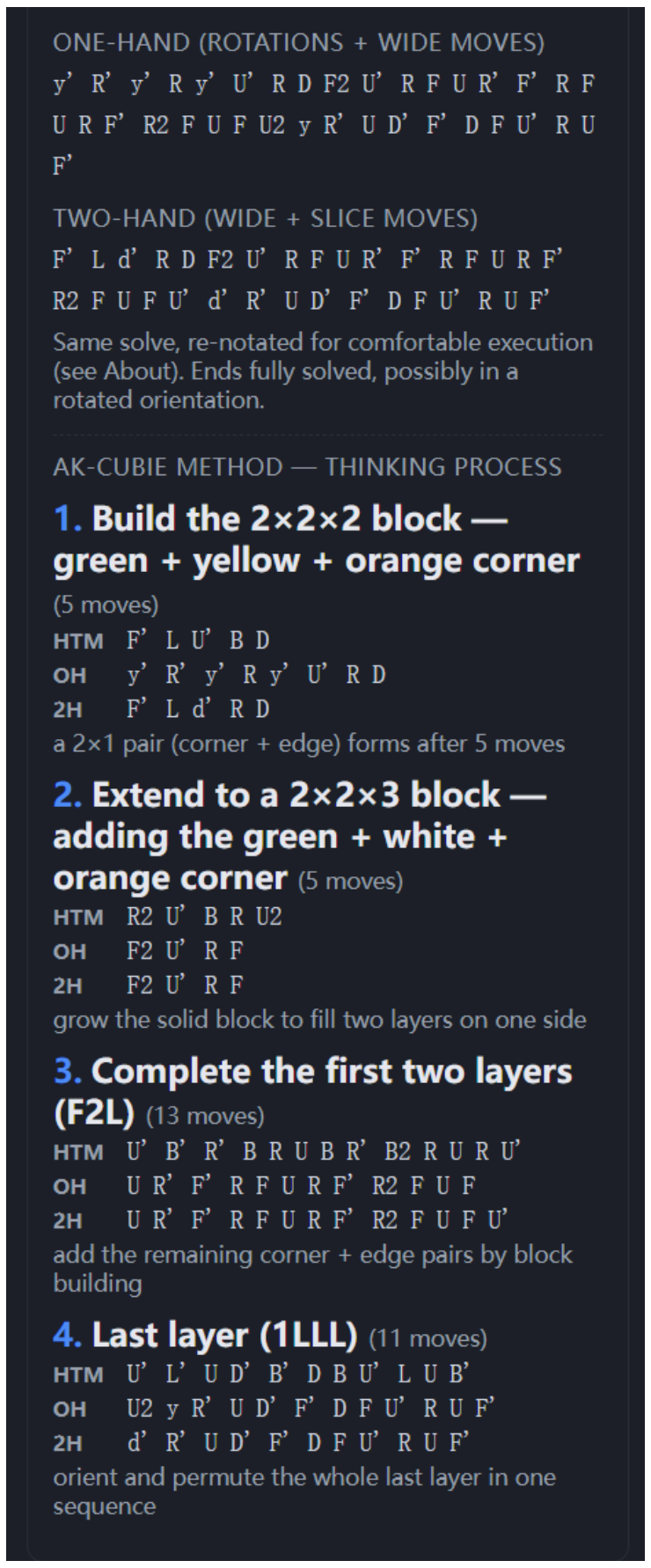

*Fig 16. AK-cubie Method*

## V. Future Development

The AK-cubie method generalizes naturally to cubes of larger sizes. Block building, or the XXX-cross, remains a key limiting factor in speedcubing, and several optimizations can be integrated into the method — including parallel processing, branch-depth control, advanced search strategies, and AI-driven heuristics. These constitute promising directions for future development.

## VI. Conclusion

The AK-cubie method can systematically construct effective, human-readable solutions. For advanced speedcubers, as typical turns per second (TPS) approaches 15, any solution of 30 moves or fewer has a strong likelihood of breaking the 2.76-second world record. With the AK-cubie software, we hope that speedcubers worldwide can acquire new techniques for block building and XXX-cross construction. Solving the cube within 2 seconds is achievable, and this possibility motivates the present work.

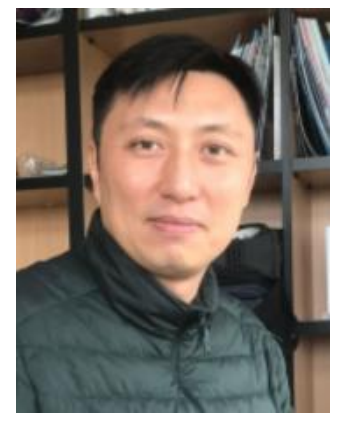

Dr. KONG, Chung To (Andy) holds a PhD in Computer Science and a Master's degree in Electrical and Electronic Engineering, both from The University of Hong Kong (HKU), as well as a Bachelor's degree in Computer Engineering from the University of British Columbia (UBC). He has accumulated extensive industry experience across diverse fields, including robotics, LED displays, FPGA, access control/CCTV systems, networking, and AI solutions. He is currently affiliated with the Laboratory for Space Research at HKU.